\documentclass[twocolumn,superscriptaddress,showpacs,oneside,prl]{revtex4}
\usepackage{amssymb}

\usepackage{epsfig,amssymb,amsfonts}

\newcommand{\be}{\begin{equation}}
\newcommand{\ee}{\end{equation}}
\newcommand{\bea}{\begin{eqnarray}}
\newcommand{\eea}{\end{eqnarray}}

\begin{document}

\title{Scalable solid-state quantum computation in decoherence-free
subspaces with trapped ions}

\author{Li-Xiang Cen}
\affiliation{ Department of Physics, Sichuan University, Chengdu
610065, China} \affiliation{Department of Physics \& Center of
Theoretical and Computational Physics, University of Hong Kong,
Pokfulam Road, Hong Kong, China}
\author{Z. D. Wang}
\email{zwang@hkucc.hku.hk} \affiliation{Department of Physics \&
Center of Theoretical and Computational Physics, University of Hong
Kong, Pokfulam Road, Hong Kong, China} \affiliation{National
Laboratory of Solid State Microstructures, Nanjing University,
Nanjing, China}
\author{S. J. Wang}
\affiliation{ Department of Physics, Sichuan University, Chengdu
610065, China}

\begin{abstract}
We propose a decoherence-free subspaces (DFS) scheme to realize
scalable quantum computation with trapped ions.  The spin-dependent
Coulomb interaction is exploited, and the universal set of
unconventional geometric quantum gates is achieved in encoded
subspaces that are immune from decoherence by collective dephasing.
The scalability of the scheme for the ion array system is
demonstrated, either by an adiabatic way of switching on and off the
interactions, or by a fast gate scheme with comprehensive DFS
encoding and noise decoupling techniques.
\end{abstract}


\maketitle

The practical accomplishment of quantum computation requires
accurate control of quantum coherent evolution to perform the
information storage for quantum bits (qubits), the processing of
information by quantum gates, and a means of final readout
\cite{QC}. The quantum computer model based on ion trap systems,
which encodes and manipulates information via long-lived internal
states of ions, was identified as one promising candidature proposal
and has witnessed rapid development in the past decade \cite
{cirac,monroe,molmer,milburn,knight,push,kiel,liebfried,garcia,duan,zhu}.
In recent literatures \cite{push,garcia,duan,zhu}, it was suggested
that quantum gates could be realized via certain spin-dependent
Coulomb interactions. These theoretical scenarios relax
significantly the physical constraints to execute gate operations,
hence offer a robust manner to implement quantum information
processing.

The scalability of the model, i.e. the extension of quantum
processors from two qubits to large numbers of ion units, however,
is quite challenging due to the growing complexity of the ion
vibrational mode spectrum. An alternative way to achieve the
scalability, which is pursued by many current efforts, is to devise
sophisticated microtrap architecture and design a series of reliable
ion shuttling \cite{push,kiel}. The scaling scenario concerning a
large array of ion crystals without ion shuttling has also been
suggested in recent proposals \cite{duan,zhu}, where a prerequisite
that the gate interactions should be comparable with the local ion
oscillation frequency is indicated.

In this paper we propose a scheme to realize scalable ion trap
quantum computation in decoherence-free subspaces (DFS)~\cite{EAC}
with an extended unconventional geometric scenario~\cite{zhuw},
which possesses the advantages of the DFS and geometric strategies:
the former is immune from decoherence induced by collective
dephasing while the latter is thought to be insensitive to certain
random errors in the operation process. We exploit the
spin-dependent Coulomb interactions and construct a universal set of
unconventional geometric quantum gates in encoded subspaces. The
potential to scale up the ion array system for quantum computation
without ion shuttling is further investigated. Two different
interaction configurations, including an adiabatic way of switching
on and off the interactions and a scenario to execute rapidly the
interaction pulses combined with a noise cancelation technique, are
presented.

The system we employed consists of $N$ trapped ions arrayed by a
convenient structure. Two stationary internal states of each ion,
denoted as $|0\rangle $ and $|1\rangle $, are selected to represent
physical qubits. We assume that the ions are separated by some
suitable distance so that on one hand the ions could be addressed
individually and on the other hand the mutual Coulomb interactions
of ions could be taken into account. In the absence of external
forces, the potential $V$ of the system is normally approximated by
the second-order expansion (the harmonic approximation) for small
vibrations around the equilibrium configuration $(q_1^{(0)},\cdots
,q_N^{(0)})$. The motional degrees of freedom of the ions is
therefore treated collectively, and the Hamiltonian in normal
coordinates reads
\begin{equation}
H_{vib}=\sum_k\frac 1{2m}P_k^2+\frac m2\sum_k\omega _k^2Q_k^2.
\label{vibh}
\end{equation}
Note that the normal coordinates relate to the local one by
$Q_k=\sum_jD_{jk}q_j$, where $D$ is an orthogonal matrix that
diagonalizes the Hessian $v_{ij}=(\frac{\partial ^2V}{\partial
q_i\partial q_j} )(q_1^{(0)},\cdots ,q_N^{(0)})$, and $\omega _k$
$(k=1,\cdots ,N)$ account for the characteristic frequencies of the
normal modes. Suppose, in order to realize the qubit-qubit coupling,
that two of the ions located at $q_i$ and $ q_j$ are exerted by
certain acceleration forces that are dependent of ion internal
states by $F_{\mu \alpha }(t)=f_\mu (t)\sigma _\alpha ^{(\mu )}$
with $\mu =i,j$. The related new interaction term hence takes a form
$ H_F(t)=-\sum_{\mu =i,j}f_\mu (t)q_\mu \sigma _\alpha ^{(\mu )}$.
We have assumed a general dependence of the forces on the qubit
states, where $ \sigma _\alpha ^{(\mu )}$ with $\alpha =x,y,z$
represent the three Pauli operators acting on the states $|0\rangle
$ and $|1\rangle $ of the qubit at site $\mu $. It is readily seen
that, by introducing the Fock operators $a_k=(m\omega
_kQ_k+iP_k)/\sqrt{2m\omega _k}$ with $[a_k,a_k^{\dagger }]=1$, the
Hamiltonian in the rotation picture with respect to $H_{vib}$ takes
the following form (setting $\hbar =1$)
\begin{equation}
H(t)=-\sum_{k=1}^N[g_i^k(t)\sigma _\alpha ^{(i)}+g_j^k(t)\sigma
_\alpha ^{(j)}]a_ke^{-i\omega _kt}+h.c.,  \label{hamil1}
\end{equation}
where $g_\mu ^k(t)=\tilde{D}_{\mu k}f_\mu (t)$ for $\mu =i,j$ and
$\tilde{D} _{\mu k}=D_{\mu k}/\sqrt{2m\omega _k}$.

To investigate the internal state evolution of the ions, we employ a
gauged representation with respect to the unitary transformation
$G(t)=\exp [-i\int_0^tH(\tau )d\tau ]$. The evolution operator of
the system is given by $U(t)=G(t)U_g(t)$, where $U_g(t)$ satisfies
the covariant equation $i\partial _tU_g(t)=H^g(t)U_g(t)$ and the
gauged Hamiltonian $H^g(t)$ is obtained as \cite{wangsj}
\begin{eqnarray}
H^g(t) &=&G^{-1}HG-iG^{-1}\partial G/\partial t  \nonumber \\
&=&\sum_{k=1}^NJ_{ij}^k(t)\sigma _\alpha ^{(i)}\sigma _\alpha ^{(j)}
+ \epsilon _0(t), \label{gaugeh}
\end{eqnarray}
where
\begin{equation}
J_{ij}^k(t)=\int_0^t[g_i^k(t)g_j^k(t^{\prime })+g_i^k(t^{\prime
})g_j^k(t)]\sin \omega _k(t^{\prime }-t)dt^{\prime }, \label{paramj}
\end{equation}
and $\epsilon _0(t)$ is merely a c-number parameter. The equation
(\ref{gaugeh}) manifests explicitly the ``spin-spin'' couplings
between the ion qubits $i$ and $j$. Especially, as the external
force is controlled with a particular configuration such that
$\int_0^TH(t)dt=0$, the transformation $G(T)$ becomes an identity
operator and the evolution operator of the system at time $T$ is
exactly the one $U_g(T)=e^{-i\Phi (T)\sigma _\alpha ^{(i)}\sigma
_\alpha ^{(j)}}=e^{-i(\Phi (T)/2)[(\sigma _\alpha ^{(i)}+\sigma
_\alpha ^{(j)})^2-2]}$ with
$\Phi (T)=\sum_{k=1}^N\int_0^TJ_{ij}^k(t)dt$.  
A key observation is  that the generated transformation $U_g(T)$
contains only Pauli operators acting on the ion qubits, therefore
the induced qubit operation is irrelevant with the ion motion degree
of freedom, hence is insensitive to the vibrational temperature of
ions. More intriguingly, similar to the case addressed in
Ref.\cite{zhuw}, the present $U_g(T)$-gate may be viewed as an
extended version of the unconventional geometric operation
\cite{zhuw} whose advantages have been demonstrated in literatures
\cite{molmer,liebfried,garcia,sackett,cenp}. This sort of gate will
be utilized as a basic one to construct the universal set of gate
operations for quantum computation in DFS.

\begin{figure}[tbp]
\begin{center}
\epsfig{figure=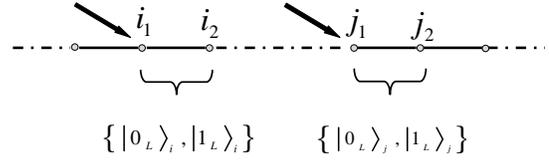,width=0.4\textwidth}
\end{center}
\caption{Schematic of encoded logical qubits for scalable ion trap
quantum computation in decoherence-free subspaces.}
\end{figure}

We employ the pair-bit code by which the logical qubit is encoded in
a subspace $C_i^2$ as
\begin{equation}
|0_L\rangle _i=|0\rangle _{i_1}\otimes |1\rangle
_{i_2},~~|1_L\rangle _i=|1\rangle _{i_1}\otimes |0\rangle _{i_2},
\label{enqubit}
\end{equation}
where $i=1,\cdots ,N/2$ indexes logical qubits and the schematic
array of ions is shown as in Fig. 1. Such an encoding constitutes
the well-known DFS \cite{EAC} against the collective dephasing of
the system-bath interaction $ \sum_{i=1}^{N/2}Z_i\otimes B$, where
$Z_i=\sigma _z^{(i_1)}+\sigma _z^{(i_2)} $ and $B$ is an arbitrary
bath operator. Let us denote  $\pi _x= $(\negthinspace {\tiny $
\begin{array}{ll}
0 & 1 \\
1 & 0
\end{array}
\!\!$}), $\pi _y=$ ({\tiny $\!
\begin{array}{ll}
0 & -i \\
i & 0
\end{array}
\!\!$}) and $\pi _z=$({\tiny $\!
\begin{array}{ll}
1 & 0 \\
0 & -1
\end{array}
\!\!$}) as the three Pauli operators of the encoded logical qubit $
\{|0_L\rangle ,|1_L\rangle \}$. A logical controlled-phase flip on
two encoded qubits $i$ and $j$, $e^{i\phi \pi _z^{(i)}\pi
_z^{(j)}}$, could be generated via aforementioned spin-dependent
Coulomb interactions as below. We exploit the acceleration forces of
the form $F_\mu (t)=f_\mu (t)\sigma _z^{(\mu )}$, where the objects
could be the two ions $\{\mu =i_1,j_1\}$, or alternatively the ions
$\{\mu ^{\prime }=i_2,j_2\}$ with the similar force configuration.
According to the previous analysis, as the force configuration is
designed so that at time $T$ there are
\begin{equation}
\eta _\mu ^k(T)\equiv \int_0^Tg_\mu ^k(t)e^{-i\omega
_kt}dt=0,~~k=1,\cdots ,N,  \label{relation}
\end{equation}
the specified interactions of Eq. (\ref{hamil1}) shall generate the
transformation $e^{-i\Phi (T)\sigma _z^{(i_1)}\sigma _z^{(j_1)}}$ or
$ e^{-i\Phi (T)\sigma _z^{(i_2)}\sigma _z^{(j_2)}}$, corresponding
to the two different addressing of forces on ions $\mu $ or $\mu
^{\prime }$, respectively. Note that the evolution generated by
these interactions falls entirely into the encoded subspace
$C_i^2\otimes C_j^2$ all along the time. Moreover owing to the
simple fact of $Z_i=Z_j=0$, the actions of both the quadratic
operators $\sigma _z^{(i_1)}\sigma _z^{(j_1)}$ and $\sigma
_z^{(i_2)}\sigma _z^{(j_2)}$ in this restricted DFS are equivalent
to that of $\pi _z^{(i)}\pi _z^{(j)}$. Therefore the gate operation
$e^{i\phi \pi _z^{(i)}\pi _z^{(j)}}$ could be exactly achieved by
any of the two interacting processes described above.

The commensurability relation (\ref{relation}), which involves all
of the ion oscillation modes, might be reached for large $N$ systems
only by an adiabatic manner to carry out the pushing forces on the
ions. In detail, let us assume that $f_\mu (t)$ characterizing the
configuration of acceleration forces is some smooth function of time
and satisfies $|\dot{f}_\mu (t)|\ll \omega _k$. Moreover, we take
$f_\mu (t)$ undergoing from $f_\mu (0)=0$ and after some finite
value to an end point $f_\mu (T)=0$. It is then direct to see that
there is $\int_0^Tf_\mu (t)e^{-i\omega _kt}dt=0$, therefore the
relations $\eta _\mu ^k(T)=0$ come into existence for all of the
oscillation modes $k$. The phase of gate operation generated in this
case has a simple form $\Phi (T)=-\sum_k(2/\omega
_k)\int_0^Tg_i(t)g_j(t)dt$. Note that the currently proposed DFS
scenario to implement qubit operations actually has already tackled
partially the intrinsic obstacle associated with the adiabatic
action, that the use of slow gates gives decoherence more time to
exert its detrimental effects.

We would also like to give some remarks on the combination of our
DFS implementation of gate operations with another scalable
approach, i.e., the fast gate scenario by using noise cancelation
techniques. In Ref. \cite{duan} it was proposed that if the
operation speed is comparable with the local ion oscillation
frequency, the noise influence due to the complexity of phonon modes
could be significantly reduced by designing a multi-cycle
configuration of kicking forces, hence the indicated scalability has
been effectively demonstrated. From our formalism, as the
commensurability relation (\ref{relation}) is spoiled, the
transformation $G_\mu (T)$ (or $G_{\mu ^{\prime }}(T)$ corresponding
to the different ion addressing) contained in the evolution operator
indicates actually a noise contribution, say,
\begin{equation}
G_\mu (T)=\exp \{i\sum_k\sum_{\mu =\{i_1,j_1\}}[\eta _\mu
^k(T)a_k+h.c.]\sigma _z^{(\mu )}\}.  \label{noise}
\end{equation}
It is seen that such an undesirable influence could be suppressed by
using two cycles of force pulses with reversal configuration. The
reason is that the induced noise effects for the two opposite
evolution will encounteract each other by viewing that the
coefficients $\eta _\mu ^k(T)$ and $\eta _{\mu ^{\prime }}^k(T)$ of
the noise operators $G_\mu (T)$ and $G_{\mu ^{\prime }}(T)$
associated with two circuits have reversed signs. Besides the direct
extension utilizing the noise cancelation via reversed loop
evolution, a slightly different approach to remove the noise effects
could be achieved for our DFS gate scenario via combining the above
specified two interactions with similar force configuration but
different ion addressing on $\mu $ and $\mu ^{\prime }$. Note the
fact that in the encoded subspace of $C_i^2\otimes C_j^2$, there are
$\sigma _z^{(i_1)}=-\sigma _z^{(i_2)}$ and $\sigma
_z^{(j_1)}=-\sigma _z^{(j_2)}$. The validity of the new approach
relies also on the assumption that the ion array system possesses a
periodic structure so that the relations
$g_{i_1,j_1}^k(t)=g_{i_2,j_2}^k(t)$ exist for the specified
interactions regarding the translation invariance of the system.
Since the requirement of the force configuration is relaxed, this
method actually provides an alternative way to implement the gate
operation for large-scale ion array systems.

To obtain fully the ability to perform quantum computation, one
needs also construct the general rotation operations for single
qubit units. Note that in the DFS schemes, nontrivial couplings
between physical qubits are necessary even to build the gates for
single logical qubits. We propose below a similar  scheme to realize
the two universal non-commuting single-qubit gates $\{e^{i\phi \pi
_x},e^{i\phi ^\prime \pi _y}\}$ in the specified DFS by utilizing
spin-dependent interactions. In detail, we make use of two different
forces $F_{\mu \alpha }(t)$ and $F_{\mu \alpha }^{\prime }(t)$ $(\mu
=1,2)$ to derive the two gates $e^{i\phi \pi _x}$ and $e^{i\phi \pi
_y}$, by which $ F_{\mu \alpha }(t)=f_\mu (t)\sigma _x^{(\mu )}$ for
the former and $F_{\mu \alpha }^{\prime }(t)=f_\mu ^{\prime
}(t)\sigma _\alpha ^{(\mu )}$ with $ \{\sigma _\alpha ^{(1)}=\sigma
_x^{(1)},\sigma _\alpha ^{(2)}=\sigma _y^{(2)}\}$ for the latter,
respectively. The corresponding interactions of the form
(\ref{hamil1}) should generate the transformations
\begin{equation}
U_F(T)=e^{-i\Phi (T)\sigma _x^{(1)}\sigma _x^{(2)}},~~U_{F^{\prime
}}(T)=e^{-i\Phi ^{\prime }(T)\sigma _x^{(1)}\sigma _y^{(2)}}
\label{gate1}
\end{equation}
at time $T$, respectively, provided that the relations
(\ref{relation}) with $\mu =1,2$ are satisfied.
Note that in the restricted subspace $\{|0_L\rangle ,|1_L\rangle
\}$, actions of the quadratic operators $\sigma _x^{(1)}\sigma
_x^{(2)}$ and $\sigma _x^{(1)}\sigma _y^{(2)}$ are exactly the same
as those of $\pi _x$ and $\pi _y$. Therefore the transformations of
(\ref{gate1}) offer actually the two gate operations $e^{i\phi \pi
_x}$ and $ e^{i\phi ^\prime \pi _y}$, respectively.

Protection of the state leakage throughout the gating period,
however, needs to be scrutinized more carefully. It is recognized
that the generated evolution of $U_F(t)$ and $U_{F^{\prime }}(t)$
actually employs the ion levels out of the DFS, the predicted
protection against collective dephasing might be spoiled during the
gate operations. For convenience, let us consider a simplified model
with only one mode with frequency $\omega $ involved, which accounts
physically for sideband addressing by laser beams to select out the
particular phonon mode. We assume further a homogeneous dependence
of the interactions on the ion internal states, that is, the
parameters in that of (\ref{hamil1}) satisfying
$g_1(t)=g_2(t)=g(t)$. Note that the evolution of the system now is
governed by the overall Hamiltonian
\begin{equation}
H_{tot}(t)=H(t)+Z_i\otimes B,  \label{totalh}
\end{equation}
where $H(t)$ is assumed to be interactions associated with $F_\mu
(t)$ or $ F_\mu ^{\prime }(t)$ with specified parameters
accordingly. It is readily seen that in the formerly described
gauged representation with respect to $ G(t)$, one obtains
\begin{equation}
H_{tot}^g(t)=H^g(t)+\tilde{Z}_i(t)\otimes B,  \label{totalhg}
\end{equation}
where $\tilde{Z}_i(t)=G^{-1}(t)Z_iG(t)$ and $H^g(t)$ has a form of
Eq. (\ref {gaugeh}) with a degenerated parameter
\begin{equation}
J(t)=2\int_0^tg(t)g(t^{\prime })\sin \omega (t^{\prime }-t
)dt^{\prime } . \label{paramjd}
\end{equation}
In detail, corresponding to the two interactions associated with
$F_\mu (t)$ and $F_\mu ^{\prime }(t)$ respectively, the form of
$\tilde{Z}_i(t)$ could be obtained as
\begin{eqnarray}
\tilde{Z}_i(t) &=&\cos \hat{\eta}^a(t)Z_i+\sin
\hat{\eta}^a(t)(\sigma
_{1y}+\sigma _{2y}),  \nonumber \\
\tilde{Z}_i^{\prime }(t) &=&\cos \hat{\eta}^a(t)Z_i+\sin \hat{\eta}
^a(t)(\sigma _{1y}+\sigma _{2x}),  \label{dephas}
\end{eqnarray}
where the operator $\hat{\eta}^a(t)=\eta (t)a+\eta ^{*}(t)a^{\dagger
}$, and
$\eta (t)=\int_0^tg(\tau )e^{-i\omega \tau }d\tau .  \label{yita}$
It is evident to see that, even for the evolution with perfect
parameter controls with $\eta (T)=0$, the occurrence of the final
terms in the expression (\ref{dephas}) for $\tilde{Z}_i(t)$ and
$\tilde{Z}_i^{\prime }(t)$ should inevitably mix the system and bath
degrees of freedom, therefore spoil the desired gate operations.

Notably, it happens that the decoherence effects induced above could
be pined down effectively via a decoupling process by devising a
symmetrized multi-circuit evolution. For instance, the first-order
of decoupling could be achieved via a two-cycle refocused
performance of the interactions indicated by $\eta (t+T_2/2)=-\eta
(t)$, where $T_2$ denotes the whole time period of the two cycles.
In view of the relations $ \int_0^{T_2}\eta ^n(t)dt=0$ with $n$ any
odd numbers and the resulted one $ \int_0^{T_2}\sin
\hat{\eta}^a(t)dt=0$, one obtains readily, by using the Magnus
expansion
\begin{equation}
U_g^{tot}(T_2)=\hat{T}\exp
\{-i\int_0^{T_2}H_{tot}^g(t)dt\}=e^{-i(h_1+h_2+\cdots )T_2},
\label{totug}
\end{equation}
the first order of the evolution with
\begin{eqnarray}
h_1 &=&\frac 1{T_2}\int_0^{T_2}H_{tot}^g(t)dt  \nonumber \\
&=&\frac 1{T_2}\int_0^{T_2}H^g(t)dt+\int_0^{T_2}\cos \hat{\eta}
^a(t)dtZ_i\otimes B.  \label{firstor}
\end{eqnarray}
The last term of Eq. (\ref{firstor}) actually contributes nothing in
our DFS systems, therefore the decoherence effects have been
effectively removed by the above  first-order decoupling process.
Physically, for the two cycles with reversed interactions, the ions
are pushed to the reverse direction, and the unwanted coupling of
the qubits with the vibrational degree of freedom induced by the
dissipation has a reverse sign. Due to this sign reversal, the
decoherence effects from these two cycles encounteract each other.
The above decoupling process could be achieved to arbitrary high
orders by iteratively application of the multiple refocusing cycles
\cite{cenp}.

The expected extension of the above operation scheme with the
comprehensive DFS encoding and noise decoupling technique to the
scalable system is argued as follows. For the scenario with fast
execution of interaction pulses, the refocusing concept for noises
cancelation for the dissipative effects and for decoherence induced
by phonon complexity is actually consistent. That is, we have shown
that the noise cancelation by decoupling process is able to remove
both the two kind of noises provided that the time scale of the
interaction pulse is fast enough comparable with the noise
frequencies. On the other hand, for the adiabatic pushing scheme,
the validity of the extension requires that $\dot{f}(t)\ll \omega
_l$ and $\dot{f}(t)\gtrsim \tau _{rel}^{-1}$, where $\omega _l$
stands for the frequency of the longest wave-length phonon mode and
$\tau _{rel}^{-1}$ denotes the relaxation rate of the internal state
of physical ions.

Before concluding, we would like to remark some features of the
present
scheme for a potential experimental implementation.
Firstly, we point out that the initialization of the logical
registers on $|0_L\rangle _i$ could be readily accomplished through
laser light addressing and manipulating individually the ions $i_2$,
i.e., initializing the states on $|1\rangle _{i_2}$. Moreover, our
scheme possesses the following advantages: (i) The noise effects of
collective dephasing, which is reported as a major source of
decoherence in the ion trap system \cite{kiel2}, could be tackled in
the scheme. (ii) Since the ion shuttling is not necessary in the
scheme, one can design the ion array in any convenient geometry with
periodic structure. (iii) There is no cooling requirement since the
scheme is insensitive to the vibrational temperature of the ions. As
has been shown, all of the oscillation modes would \textit{not}
spoil the global operation generated by an adiabatic manner of the
evolution. For the fast gate scenario, by making use of the noise
cancelation technique, there is actually very weak influence of the
vibrational temperature on the gate fidelity \cite{duan}.

In summary, we have proposed a DFS scheme to implement scalable ion
trap quantum computation with an extended unconventional geometric
approach. Using the spin-dependent Coulomb interactions, we show
that the universal set of quantum gates could be achieved in DFS,
either via the adiabatic manner to switch on and off the
interactions, or via the scenario to execute rapidly the interaction
pulses combined with noise cancelation techniques.

This work was supported by the RGC grant of Hong Kong (HKU7045/05P),
the URC fund of HKU , and the NSFC grants (10375039, 10429401, and
90503008).

\end{document}